\documentclass[12pt]{article}
\usepackage{graphicx}

\usepackage{amsmath,amssymb,graphicx,multirow,xspace}
\usepackage[colorlinks=true,urlcolor=blue,anchorcolor=blue,citecolor=blue,filecolor=blue,linkcolor=blue,menucolor=blue,pagecolor=blue]{hyperref}
\usepackage{subfigure, placeins}
\usepackage{booktabs}
\usepackage{blindtext, rotating}
\usepackage{afterpage}
\usepackage{enumitem}
\usepackage{ marvosym }
\usepackage{verbatim}
\usepackage{authblk}

\newcommand\pubnumber{}
\newcommand\pubdate{\today}

\def\uiuc{Department of Physics, University of Illinois at Urbana-Champaign\\ Urbana, IL 61801}

\def\Title#1{\begin{center} {\Large #1 } \end{center}}
\def\Author#1{\begin{center}{ \sc #1} \end{center}}
\def\Address#1{\begin{center}{ \it #1} \end{center}}

\newcommand\pubblock{\rightline{\begin{tabular}{l} \pubnumber\\
         \pubdate  \end{tabular}}}
\newenvironment{Abstract}{\begin{quotation}  }{\end{quotation}}
\newenvironment{Presented}{\begin{quotation} \begin{center} 
             PRESENTED AT\end{center}\bigskip 
      \begin{center}\begin{large}}{\end{large}\end{center} \end{quotation}}
\def\Acknowledgements{\bigskip  \bigskip \begin{center} \begin{large}
             \bf ACKNOWLEDGEMENTS \end{large}\end{center}}

\def\beq{\begin{equation}}
\def\eeq#1{\label{#1}\end{equation}}
\def\eeqn{\end{equation}}

\def\beqa{\begin{eqnarray}}
\def\eeqa#1{\label{#1}\end{eqnarray}}
\def\eeqan{\end{eqnarray}}

\def\st{\scriptstyle}

\let\bar=\overbar

\def\Dslash{\not{\hbox{\kern-4pt $D$}}}
\def\dslash{\not{\hbox{\kern-2pt $\del$}}}

\def\msb{{\bar{\ssstyle M \kern -1pt S}}}

\newcommand{\bea}{\begin{eqnarray}}
\newcommand{\eea}{\end{eqnarray}}
\newcommand{\mat}{\begin{pmatrix}}
\newcommand{\rix}{\end{pmatrix}}

\renewcommand{\bar}{\overline}
\renewcommand{\slash}[1]{#1\!\!\!/}

\newcommand{\go}{{\tilde g}}

\newcommand{\Ho}{{\tilde H}}

\renewcommand{\st}{{\tilde t}}

\newcommand{\stau}{{\tilde\tau}}

\renewcommand{\beqa}{\begin{eqnarray}}
\renewcommand{\eeqa}{\end{eqnarray}}

\renewcommand{\beq}{\begin{equation}}
\renewcommand{\eeq}{\end{equation}}

\newcommand{\order}[1]{{\cal O}\left(#1\right)}

\newcommand{\met}{{\slash E_T}}

\begin{document}
\begin{titlepage}
\pubblock

\vfill
\Title{Flavors of Supersymmetry Beyond Vanilla}
\vfill
\Author{ Jared A.~Evans} 
\Address{\uiuc}
\vfill
\begin{Abstract}
This review surveys the territory of supersymmetry beyond the vanilla MSSM.  With a viewpoint guided by electroweak naturalness, the review focuses on constructions that weaken or bypass current LHC constraints.  Models of SUSY containing Dirac gluinos, compressed spectra, flavor-violating squarks, R-parity violation, stealth sectors, exotic detector objects, and more are discussed.  In addition to presenting ways of hiding SUSY, these models highlight a few opportunities to improve LHC coverage.
\end{Abstract}
\vfill
\begin{Presented}
Conference on the Intersections of Particle and Nuclear Physics   \\ \vspace{2mm} CIPANP2015 -- Vail, CO, USA, May 19\,--\,24, 2015
\end{Presented}
\vfill
\end{titlepage}
\def\thefootnote{\fnsymbol{footnote}}
\setcounter{footnote}{0}

\section{Introduction}

 There are many alluring properties of low-energy supersymmetry (for a nice, general introduction, see \cite{Martin:1997ns}).  It is a viable space-time extension,  provides a natural solution to the hierarchy problem, facilitates gauge coupling unification at a high scale, and, with a stable, neutral LSP (lightest supersymmetric particle), provides a realistic WIMP dark matter candidate.  However, despite the numerous appealing aspects, low-energy supersymmetry (SUSY) is plagued by one overwhelming failure: we have seen no compelling evidence for it at the LHC \cite{Craig:2013cxa}.  The apparent absence of light superpartners puts supersymmetry in serious tension with electroweak naturalness.  Although this failure may tempt some to abandon low-energy SUSY, one could instead consider the possibility that SUSY does not behave as the vanilla MSSM variety that was expected to appear at the LHC.  It instead could manifest in some way that allows it to evade the ever rising constraints, but to preserve some or all of these desirable features that initially drew us to SUSY.  
 
\begin{figure}[htb]
\centering
\includegraphics[height=1.5in]{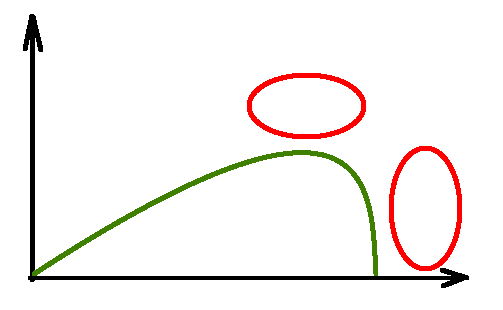} \hspace{1cm} \includegraphics[height=1.5in]{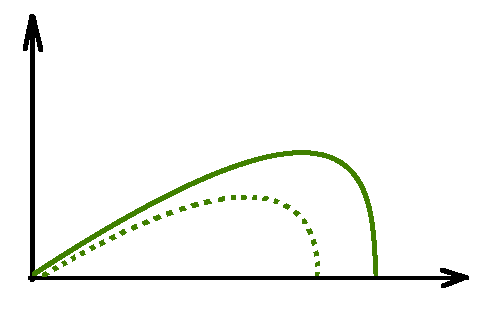}
\put(-62,-1){$m_{\mbox{\scriptsize production}}$}
\put(-195,80){$m_{\mbox{\scriptsize LSP}}$}
\put(-274,-1){$m_{\mbox{\scriptsize production}}$}
\put(-408,80){$m_{\mbox{\scriptsize LSP}}$}
\caption{{\bf Left:} Heavy production state or LSP that lives ``out of reach'' beyond the green exclusion curve, e.g., in the red circles. {\bf Right:}  Degraded efficiency through competing branching ratios or modified kinematics weaken limits, taking the solid exclusion curve to the dashed one. }
\label{fig:behavior}
\end{figure}

 What exactly is vanilla SUSY?  For the purpose of this review, vanilla SUSY will mean what the typical SUSY searches are looking for, i.e., an accessible colored state (the production particle) which cascade decays through a restricted chain to a stable, neutral LSP with a mass significantly lower than the production particle.  If we want to believe in low-scale SUSY, there are three basic options:
 \begin{enumerate}
 \item  Superpartners are beyond the exclusion curves of existing searches
 \item  Superpartners are lighter, but manifest with a reduced efficiency to populate searches
 \item  Superpartners decay in a way to which existing searches are insensitive
 \end{enumerate}
 The effects of the first two options on existing limit plots are schematically illustrated in Figure~\ref{fig:behavior}.  In essence, each of these three options is doing one of two things:  reducing the signal below the point where it can be seen over the SM background or, the more exciting possibility, moving the model into a region of  signature space currently unexplored by LHC searches.

 The purpose of this review is to discuss the viability of many of these options with a focus on recent progress.  Although I depart from this assumption in many places, my view will be tempered by a natural electroweak sector, which is the main reason to expect supersymmetry will be LHC accessible.  I will not discuss models of SUSY that accommodate the observed Higgs mass, which tend to drastically increase the amount of tuning required.  I will first discuss the possibility of pushing superpartners beyond the scale viewed as natural.  I will then discuss methods to degrade efficiency, either through use of modified kinematics or competing branching ratios.  Lastly, I discuss some methods whereby SUSY can exist in way that would avoid the existing LHC constraints, either by removing $\met$ or introducing exotic detector objects.  As the scope is fairly broad, this review is not intended to be comprehensive, but rather it is aims to provide a sample of possibilities.

\section{Out of Reach}

  In order to get a light electroweak scale from weakly coupled SUSY without requiring a finely tuned cancellation between bare parameters and quantum corrections, the model requires light gluinos, stops and Higgsinos (see e.g., \cite{Papucci:2011wy}).  Tuning is a qualitative concept for which quantitative measures can be constructed, thus ``tuning constraints'' are both pliable in definition and a matter of personal preference.  Roughly speaking, requiring no more than 20\% (10\%) tuning in the MSSM requires that Higgsinos, both stops, and the gluino are less than about 200, 500, and 900 GeV (350, 700, and 1500 GeV), respectively.  LHC limits on these particles have begun to push these boundaries.  One apparent option to raise the mass of the superpartners is to simply ``pay the tuning cost,'' although it is unclear exactly what this means without appealing to an anthropic landscape.  Briefly, I will discuss the constraints and some of the theoretical options that can permit living beyond the reach of existing searches in a natural framework.
  
 \subsection{Gluino Production}

  As the gluino production rate is so staggeringly large at the LHC, these particles cannot be much less than 1 TeV independent of their decay path \cite{Evans:2013jna} (barring a few oddball exceptions discussed later in this review).  Depending on your personal tuning standards, these constraints are already at odds with a natural electroweak sector.   
  
  However, if the gluinos are Dirac particles, then their contributions to radiative corrections have no divergences, a.k.a.~they are ``supersoft'' \cite{Fox:2002bu}.  Dirac gluinos, which emerge in U$(1)_R$-symmetric models, could realistically be a few TeV and allow for a natural electroweak sector.   As Dirac gluinos do not contribute to the valence quark $t$-channel exchange diagrams, e.g.,~$uu\to \tilde u\tilde u$ production, models containing them are also ``supersafe'' \cite{Kribs:2012gx}, meaning first-generation squarks can be light and evade constraints.  The once theoretically dire issues with QCD breaking are absent in models with ``Goldstone Gauginos'' \cite{Alves:2015kia,Alves:2015bba}.  Overall, Dirac gluinos are one of the most promising options for preserving many of SUSY's theoretical glories while accounting for its absence at the LHC.  It is worth noting that light Dirac gluinos are generally \emph{more} excluded (by a factor of two) than their Majorana counterparts \cite{Evans:2013jna}.

  \subsection{Stop Production}
 Stops have a much lower production rate than gluinos, and thus receive less stringent constraints from LHC searches.  In fact, limits are weak enough that for Higgsinos near 300 GeV, there are no constraints on stops \cite{Aad:2015pfx}.  Also, if $m_\st-m_{LSP}\sim m_t$, there are almost no constraints (although, see \cite{Han:2012fw,Aad:2014mfk}).  Although a lot of natural territory still remains for vanilla stops, Run II should be able to constrain much of it (or make a discovery) \cite{ATLAS:2013hta}.  Unlike gluinos, there are no known, compelling, natural methods to separate the stops from the electroweak sector.
 
\subsection{Higgsino Production/LSP} \label{sec:higgsino}

Despite the smaller production rate of Higgsinos, there are meaningful constraints on Higgsino production, but these constraints rely on either large splittings or a superpartner even lighter than the Higgsino in the spectrum.  Direct production of near degenerate LSP Higgsinos currently has no meaningful LHC constraints.  While some space will hopefully be probed, the reach for Higgsinos with very small splittings (as arises in models with heavier $M_1$ and $M_2$) will likely remain unconstrained during Run II \cite{Han:2014kaa}.  

Although light Higgsinos are typically required for a natural electroweak scale, there are possible exceptions \cite{Cohen:2015ala}.  Constraints on gluinos/stops can be reduced by raising the Higgsinos (and thus the LSP) to give a compressed spectrum.    As the observed $\met$ is restricted by the \emph{visible} energy in the event, compressed spectra result in lower $\met$.    Once one has $\frac{m_{LSP}}{m_{production}}\gtrsim 0.3$, the reduced visible energy begins to result in a substantial loss of observed missing energy.  When one has $1 - \frac{m_{LSP}}{m_{production}}\ll 1$, i.e., the LSP is only a little lighter than the production particle, then observing any $\met$ relies on the presence of hard initial state radiation.  Interestingly, models of SUSY with extra compact dimensions can give rise to compressed spectra rather generically \cite{Murayama:2012jh}.  Projections for probing compressed spectra via gluinos at Run II are rather strong \cite{Bhattacherjee:2013wna}, and should be able to close in on this loophole.
   
  \section{Degradation of Efficiency}
  
 As an alternative to simply living outside of the current bounds, SUSY could instead have a production particle and LSP with masses within the nominally excluded regions, but manifest with a reduced efficiency to populate these constraining final states.  This can be accomplished either by increasing the number of available final states or by modifying the kinematics to more regularly fail the cuts of the specific search.

\subsection{Increasing Available Final States}

As nearly all searches for supersymmetry use either simplified models or heavily constrained parameter spaces (e.g., mSUGRA), it is possible that in an actual model the production particle could have multiple available decay paths to get to the LSP.   If a search is only sensitive to one decay path, and only fraction $x$ of the production particles decay that way, then the efficiency is effectively degraded by $x^2$.  Since a lot of territory in the excluded regions is often only excluded by $\order{1}$ factors rather than by orders of magnitude, this can serve to reduce constraints on superpartner masses by quite a bit.  Of course, doing this necessarily opens up additional decay paths which could potentially set constraints, either on their own or in conjunction with the other decay paths.

One interesting possibility is flavor-mixed squarks.  Historically, flavor-mixed squarks have been largely neglected due to potentially dangerous flavor constraints, especially meson mixing.  However, models where the squark mixing exhibits chiral flavor violation \cite{Evans:2015swa} have vastly reduced constraints, especially in the right-handed sector.   A mix of charm- and top-flavored squark has greatly reduced collider constraints \cite{Blanke:2013zxo,Agrawal:2013kha}.  Even a mix of up- and top-squark could be rather light and consistent with current observations \cite{Backovic:2015rwa}, so direct searches for these mixed signatures are of great interest.

 Perhaps the best existing example of multiple possible final states comes from the pMSSM \cite{Berger:2008cq}.  In this framework, myriad models consisting of 19 or 20 independent parameters are generated by randomly assigning values.  A recent study of $\sim 2.3\times10^5$ pMSSM models \cite{Cahill-Rowley:2014twa} found only a handful of spectra that live under the simplified model constraints, and that handful falls only slightly under the bounds.  Of course, with 19 or 20 parameters, this is obviously a very sparse scan, i.e., $2^{20}\sim 10^6$, but at least this suggests that simplified models provide a fairly representative picture, and significant degradation is not a generic feature.  It would be interesting to systematically optimize this multiple final state degradation in the MSSM to determine how low one can really go with this approach.
 
 \subsection{Modifying Kinematics}

Another option is to modify the kinematics of the decay chain.  Introducing and/or squeezing intermediate particles can soften the $p_t$ of objects.  With the spectra so arranged, these objects can consistently fall below the $p_t$ thresholds used in particular searches, and thus reduce constraints by some amount.  

As another example, one could mandate \emph{more} invisible objects occurring in the decay chain.    As discussed in section \ref{sec:higgsino}, having less visible energy can reduce the overall $\met$.  Therefore, by requiring an NLSP (next-to-lightest supersymmetric particle) to undergo a three-body decay into one visible and two invisible objects, the efficiency to fall under standard jets+$\met$ searches can be significantly reduced \cite{Alves:2013wra}.  Generically, this can be accomplished through charging the LSP under an additional $\mathbf Z_2$ symmetry.  This method is most effective when there are very few visible objects in the event.

\section{Final States Beyond Vanilla}

\subsection{No $\met$}

One of the key assumptions of nearly all SUSY searches is the presence of missing energy.  In addition to only getting missing energy through ISR in compressed spectra, there are two other well-known mechanisms which can remove $\met$ in a SUSY decay chain: $R$-parity violation \cite{Barbier:2004ez} and stealth \cite{Fan:2011yu}.  $R$-parity violating couplings need to be forbidden in the usual MSSM by imposing baryon and lepton number.  With these violated, the LSP can decay, the dark matter candidate is lost, and there are potentially dangerous precision constraints that mandate a rather hierarchical structure to the RPV couplings (small universality is also permitted by the constraints).    Stealth SUSY introduces nearly degenerate states, e.g., $\tilde S,\, S$, in a nearly supersymmetric hidden sector, so $m_{\tilde S} - m_S \ll m_S$,  which transition to a stable, but very light LSP, e.g., $\tilde S \to S \tilde G$, so that the actual $\met$ produced in the decay is less than that from jet mismeasurement.  Unfortunately, the simplest framework of decay into a gravitino produces displaced decays which receive serious constraints unless the stealth sector particles are very light (see section \ref{sec:disp}).  More complicated frameworks, such as a light axino with a low decay scale, represent a viable mechanism to generate stealthy decays.

 Removing the LSP from the event only serves to appreciably weaken constraints when the final states are either all-hadronic or heavily tau enriched \cite{Evans:2012bf}.  Decays of gluinos or stops that give rise to hard neutrinos are still quite constrained by typical $\met$-based searches \cite{Evans:2013jna,Evans:2012bf}.  Gluinos and stops that give hard leptons, but no missing energy find leptoquark searches \cite{Khachatryan:2015vaa} extremely constraining \cite{Evans:2013uwa}, even with the additional jets in the final state.  For the most part, gluinos decaying to an all-hadronic final state in either RPV or stealth SUSY receive significant constraints from the CMS black hole search \cite{Chatrchyan:2013xva} and the Atlas search for 6-7 high-$p_t$ jets \cite{Aad:2015lea}.  These two searches complement one another, and are able to constrain $\go\to n j$ decays for $n\gtrsim4$ and $n\lesssim5$, respectively \cite{Evans:2013jna}.
 
For all-hadronic stop decays, much theoretical and experimental effort has been devoted to achieving sensitivity to the $\st \to jj$ signature, wherein stops decay through an RPV $\lambda''_{312}  U_3 D_1 D_2$ operator.  With the full 8 TeV data set, CMS has been able to constrain RPV stops in the 200-350 GeV range \cite{Khachatryan:2014lpa}.  Substructure method are projected to be able to reach lower, ideally to the LEP bound \cite{Bai:2013xla}.  Run II prospects for sensitivity look very strong, with limits reaching up to $\sim750$ GeV, and up to $\sim1$ TeV with the high-luminosity data \cite{Duggan:2013yna}.

\subsubsection{Jet $p_t$ Hierarchies}
  
\begin{figure}[htb]
\centering
\includegraphics[height=1.5in]{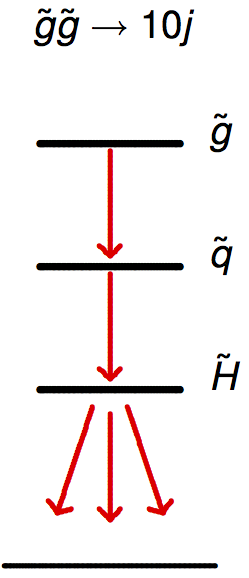} \hspace{0.5cm} \includegraphics[height=1.5in]{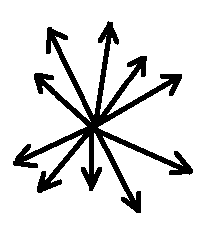} \hspace{1.5cm} \includegraphics[height=1.5in]{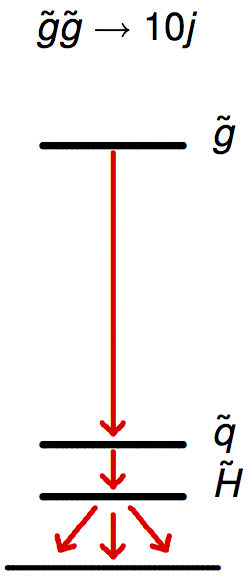} \hspace{0.5cm} \includegraphics[height=1.5in]{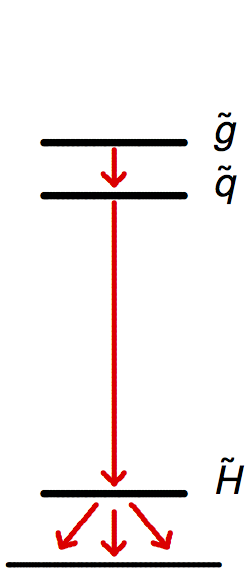} \includegraphics[height=1.5in]{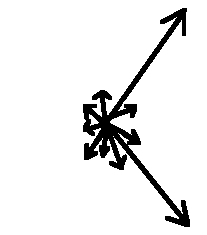}
\caption{{\bf Left:} The figure schematically illustrates a spectrum where each jet receives an approximately equal share of the system's energy.  {\bf Right:}  These two spectra exhibit a jet $p_t$ hierarchy, resulting in two very hard jets and many, softer objects that result in fewer jets measured in the actual searches.}
\label{fig:jpth}
\end{figure}

  In all-hadronic $\go$ events, one can avoid constraints from the black hole and high-$p_t$ jet searches (\cite{Chatrchyan:2013xva,Aad:2015lea}) by introducing a jet $p_t$ hierarchy \cite{Evans:2013jna}.  A jet $p_t$ hierarchy appears when there is a single large splitting in the relevant spectrum, see fig.~\ref{fig:jpth}.  This gives a signature of one hard jet from each gluino and several softer partons which routinely fail the jet selection requirements.  This behavior is more QCD-like and, thus, more difficult to probe.  Still, there may be promising handles to explore these scenarios \cite{Evans:2013jna}, but  there has been no detailed study to date.
    
\subsubsection{High $b$-jet Multiplicity}  
  
One gap in the existing search program is looking for a high multiplicity of $b$-tagged jets in all-hadronic events.\footnote{A very recent CMS paper \cite{CMS:2015wva} presents information on a single bin requiring five $b$-tagged jets in an all-hadronic search.  A single event is observed with 1 TeV $<H_T<1.75$ TeV for exactly 5 $p_{T,j}>50$ GeV jets, all of which are $b$-tagged.}  This can manifest in a variety of scenarios, both with and without SUSY, e.g.,  \cite{Evans:2014gfa,Fan:2015sza,Fox:2015kla}, but one specific, motivated SUSY scenario can be achieved with RPV.  In the MFV ansatz for RPV SUSY \cite{Csaki:2011ge}, the largest RPV operator is $\lambda''_{323} U_3 D_2 D_3$.  While in principle stops could decay to $jb$ through this operator, if the Higgsino is lighter than the stop, then the decay through  $y_t$ to $b\Ho^+$ will typically dominate.  The Higgsino itself can then undergo a three-body decay back through the off-shell stop to $jbb$, yielding a net 8-jet, 6-$b$ final state.

Unlike the background to most all-hadronic signals, the five and six $b$-jet backgrounds are expected to be very small \cite{Evans:2014gfa}.   Even in this scenario, with the rather low stop production rates, the new physics can overwhelm the background.  The presence of an internal resonance can be used to disambiguate signal events from an unexpectedly large background.
 
\subsection{Exotic Detector Objets}

Instead of a traditional decay path that involves standard physics objects, i.e.,  photons, electrons, muons, hadronic $\tau$s, jets, $b$-jets and $\met$, SUSY events originating from the production superpartner could contain an exotic object that behaves in an atypical way and defies the existing molds for object criteria.  For instance, in the case of a nearly degenerate wino LSP, the chargino can decay into the neutralino emitting a soft lepton.  If this decay has a lifetime of $c\tau\sim \order{1\mbox{ m}}$, then the signature can manifest as a disappearing track -- an object that does not appear as any of the detector objects listed above, and limits on such scenarios are much more powerful than the reach without the ``disappearing track'' exotic object \cite{CMS:2014gxa}.

\subsubsection{Displaced \& Detector Stable} \label{sec:disp}

  One of the most motivated ways to produce an exotic detector object is to have a heavy particle that propagates a macroscopic distance through the detector before decaying.  This requires an abnormally long lifetime, but small couplings can arise in a variety of motivated scenarios, such as the decays to a gravitino in GMSB, very small couplings in RPV or dynamical RPV (dRPV) \cite{Csaki:2013jza}, or the inter-gaugino transitions through very heavy squarks that arise in mini-split SUSY \cite{Arvanitaki:2012ps}.  Fortunately, most of these scenarios have been considered either implicitly or explicitly at the LHC.  The constraints on these models are very strong \cite{Liu:2015bma,Csaki:2015uza} in both the case of a heavy detector-stable charged particle \cite{Chatrchyan:2013oca} and shorter lifetime displaced decays \cite{CMS:2014wda,CMS:2014hka,Aad:2015rba}.    Much of the displaced territory is already covered by the 8 TeV program.
  
One notable gap is the case of displaced slepton decays.  An NLSP $\stau_R$ is roughly unconstrained by the CMS search for $e^\pm\mu^\mp$ with high-impact parameters  \cite{Khachatryan:2014mea} due to a variety of factors, including the small production rate and branching ratio and the right-handed polarization of the $\stau$ \cite{EvansShelton}.  13 TeV updates to the search have the potential to weakly constrain this scenario.  The case of same-flavor leptons with high-impact parameters (not originating from the same vertex), such as would arise with an NLSP $\tilde e$ or $\tilde \mu$, falls under the purview of no current search.  Such a gap is especially dangerous because of cosmic muon vetoes that exist in many new physics searches for standard objects and will reject entire events from being considered at all.

\subsubsection{Non-isolated Leptons}

  A decay chain that possesses a non-isolated lepton, i.e., a lepton that fails the isolation requirements in standard searches, has the potential to mask low-energy SUSY \cite{Graham:2014vya}.   This can arise from the decay of a very light bino ($\lesssim 10$ GeV) through a $UE\bar D$ dRPV operator \cite{Csaki:2013jza}.  While it can be challenging to create such an exotic object from a theoretical standpoint, this possible signature needs to be addressed.    A dedicated search should be able to constrain these objects even against the heavy flavor background \cite{Brust:2014gia}.

\subsubsection{Even More Exotic Exotics}

 Another way that SUSY could be hidden is if the decay chain of the production particle always passes through a hidden valley \cite{Strassler:2006im} or results in a quirk pair \cite{Kang:2008ea}.  This can create a scenario with a very odd detector signature that can be vetoed by quality criteria.  As an example, stops could decay into a hidden sector resulting in an ``emerging jet'' \cite{Schwaller:2015gea}, an exotic object where late decays from light hidden valley states produce substantial hadronic activity in the calorimeter, but exhibit reduced activity in the early layers of the tracker.  While such a scenario could easily be hidden currently, a dedicated search for such an exotic object would be able to place meaningful bounds.  
 
 In general, probing exotic detector objects involves dedicated analyses.  Determining where the existing program has deficiencies, and whether a minimal set of searches can cover all exotic object loopholes is an important goal for both theorists and experimentalists in the coming era.

\section{Conclusions}

\newcommand{\vs}{ \vspace{-2mm}}

 As LHC exclusions rise and vanilla supersymmetry grows increasingly unnatural, it is imperative that we look to models beyond this minimal variety.   In these models, we can hope to preserve some of SUSY's allure, while skirting the powerful LHC constraints.     Theoretical proposals can allow us to decouple gluinos and Higgsinos from the electroweak sector in order for these heavy states to exist beyond current limits naturally.  The LHC constraints can be relaxed by reducing the efficiency to populate final states, either by changing the branching ratios or by modifying kinematics.  In principle, by constructing models with unconventional decays that involve either no $\met$ or exotic objects, standard SUSY searches become insensitive.
 
 Fortunately, the LHC collaborations have achieved a very strong coverage of supersymmetry, not only in the vanilla case, but also across many of these more exotic possibilities.  While trying to maintain a natural electroweak sector, gaps in coverage can still be produced, but this typically requires very contrived scenarios.  The influx of 13 TeV data should fill many of these gaps, but there are a few cases where sensitivity can be greatly improved or where a dedicated search is necessary.  Some of these mentioned in this review include: \vs
\begin{itemize}
\item Flavor-violating squark decays \vs
\item Jet $p_t$ hierarchies \vs
\item High $b$-jet multiplicities \vs
\item Displaced NLSP $\stau$s/$\tilde \mu$s/$\tilde e$s \vs
\item Non-isolated leptons \vs
\end{itemize} 
 Additionally, a thorough exploration of hidden valleys, quirks and other nonstandard detector objects is necessary to probe the most exotic of SUSY flavors.  Despite the baroque constructions that produce these gaps,  it is paramount that they be filled in order to provide us comprehensive LHC coverage of new physics, in supersymmetry and beyond.

\Acknowledgements
I am grateful to Adam Martin and the CIPANP organizers for inviting me to give this talk.

\end{document}